\documentclass[12pt,preprint]{aastex}
\usepackage{psfig}
\def\HI {H\kern0.1em{\sc i}} 
\def\water {H$_{\rm 2}$O}
\def\txs {TXS\,2226{\tt -}184}
\def\pks {PKS\,2322{\tt -}123}
 
\def\etal {{\sl et~al.\ }}

\def\deg{$^{\circ}$}
\def\kms{km s$^{-1}$}

\begin{document}
\title{~~\\ ~~\\ \HI\ Absorption in the Gigamaser Galaxy TXS\,2226{\tt -}184
and the Relation between \HI\ Absorption and Water Emission}
\shorttitle{\HI in \txs}
\shortauthors{Taylor \& Peck}
\author{G. B. Taylor\altaffilmark{1}, A. B. Peck\altaffilmark{2,3}, 
C. Henkel\altaffilmark{2}, H. Falcke\altaffilmark{2}, 
C. G. Mundell\altaffilmark{4}, C. P. O'Dea\altaffilmark{5}, 
S.A. Baum\altaffilmark{5}, \& J. F. Gallimore\altaffilmark{6}}

\altaffiltext{1}{National Radio Astronomy Observatory, P. O. Box 0, Socorro, NM
  87801, USA}
\altaffiltext{2}{Max-Planck-Institut f\"ur Radioastronomie, Auf dem H\"ugel 69, D-53121 Bonn, Germany}
\altaffiltext{3}{{\it Current address:} Harvard Smithsonian Center for 
Astrophysics, SAO/SMA Project, P.O. Box 824, Hilo, HI 96721, USA}
\altaffiltext{4}{Astrophysics Research Institute, Liverpool John Moores University, Twelve Quays House,
Egerton Wharf, Birkenhead, CH41 1lD, UK}
\altaffiltext{5}{Space Telescope Science Institute, Baltimore, MD 21218, USA}
\altaffiltext{6}{Department of Physics, Bucknell University, Lewisburg, PA 17837, USA} 


\slugcomment{As Accepted by the Astrophysical Journal}

\begin{abstract}

We report on the discovery of \HI\ in absorption toward the 
gigamaser galaxy \txs\ using the VLA.  The absorption appears to consist of
two components -- one with a width of 125 \kms, and one broader 
(420 \kms), both
toward the compact radio source in the nucleus of the galaxy.
Based on these large velocity widths we suggest that the \HI\ absorption
is produced in the central parsecs of the galaxy, on a similar scale to
that which gives rise to the water maser emission.  This brings to 
eight the number of galaxies known to exhibit both water masers and
\HI\ absorption.  We explore the relationship between these two
phenomena, and present a physically motivated (but unfruitful)
search for water maser emission in five radio galaxies known to
exhibit strong \HI\ absorption.  

\end{abstract}


\section{Introduction}

A number of studies (e.g., van Gorkom \etal\ 1989, Pihlstr\"om 2001)
have shown that the rate of detection of \HI\ absorption in galaxies
is highest in Compact Symmetric Objects (CSOs) and Compact Steep
Spectrum (CSS) objects.  Of the theories proposed to explain this
phenomenon, the existence of a circumnuclear disk or torus structure
seems the most likely.  In this scenario, it is the orientation and
geometry of the source which increases the likelihood of detecting
\HI\ in absorption.  In the CSOs and CSS objects the emission is dominated by
the hot spots and lobes which are not Doppler boosted. Thus, in the
CSOs we have bright symmetric emission against which to search for \HI\ 
absorption on parsec scales at a large range of source inclination
angles.  In large sources, however, the parsec scale emission is
dominated by the relativistic jets. It is much harder to see \HI\
absorption in these sources since when the source is oriented towards
us, the counter jet is deboosted and any disk which is perpendicular
to the jet is at an unfavorable angle for absorption.


Water megamasers are found predominantly in Seyfert 2 (or obscured)
galaxies for similar reasons (Braatz, Wilson \& Henkel 1997).  The
masing gas is usually thought to be located in a circumnuclear
accretion disk ($e.g.$ NGC\,4258, Miyoshi et al.\ 1995; NGC\,1068,
Greenhill \& Gwinn 1997).  In order for the pathlength through the
disk to be long enough to obtain significant amplification, the disk
must be viewed nearly edge-on.  Neufeld \& Maloney (1995) predict that
in sources that have a molecular accretion disk sufficiently dense in
the central parsec to generate stimulated emission, atomic gas should
be present above and below the plane of this disk where the pressure
is slightly lower, as well as at larger radii.  Although this atomic
gas is expected to have a temperature as high as 8000 K, resulting in
very low optical depths, detection is possible if the continuum source
is bright enough.  Thus H$_2$O megamaser sources which have
sufficient continuum flux at 1.4 GHz are prime candidates to search
for \HI\ absorption (see NGC\,1068, Gallimore, Baum \& O'Dea 1996;
NGC\,3079, Satoh et al.\ 1997).

Several symmetric radio sources which exhibit \HI\ absorption have been
studied with the VLA and the VLBA.  A few of these are consistent with
the parsec-scale circumnuclear torus model ({\it i.e.} 1946+708, Peck,
Taylor \& Conway 1999; \pks, Taylor et al.\ 1999; Hydra~A,
Taylor 1996).  These sources exhibit very broad (several hundred \kms)
shallow systems of multiple blended lines, usually slightly redshifted
with respect to the systemic velocity of the galaxy.   Not all \HI\
absorption systems are interpreted as the result of a parsec scale
torus however.  Another class of
radio sources toward which \HI\ absorption is seen have very
distinct types of absorption systems which are also close to the
systemic velocity of the galaxy.  These systems are comprised of only
a few strong, very narrow (3--20 \kms) lines, and vary little in FWHM
or optical depth across the central parsecs of the continuum source
({\it i.e.,} Cen~A, Peck \& Taylor 1998a).
This is thought to be due to a larger, kiloparsec scale disk or dust
lane in the host galaxy.  In both cases, only a few sources have
been well studied.

Thus both \HI\ and H$_2$O can be tracers of circumnuclear material.
By combining observations of both types, we are attempting to better
constrain the distribution of material in the central parsecs of AGN.
Our survey to date encompasses several types of radio sources
(Seyfert 2, LINER, CSO, etc.) with a large range of radio powers,
central masses and estimated accretion rates.  The standard model of the
central few parsecs should scale accordingly.  It might be possible
for a radio galaxy with a central mass of 10$^9$ $M_\odot$ to have a
molecular disk of radius $\sim$20 pc and maser emission with apparent
$\sim$10$^5$ $L_\odot$ isotropic luminosity  (Neufeld \& Maloney 1995).
Here we present a search for H$_2$O maser emission towards symmetric
sources in which the \HI\ absorption is thought to arise in a
circumnuclear torus, and a search for \HI\ absorption towards the
H$_2$O gigamaser galaxy \txs.

In \S 2 we describe the search for water masers toward ``\HI\
torus'' sources carried out with the Effelsberg 100-m telescope, and
in \S 3 we describe VLA observations of \HI\ absorption toward \txs.  A
discussion of our current understanding of the circumnuclear
environment is presented in \S 4.

We assume H$_0 = 75$ km s$^{-1}$ Mpc$^{-1}$ and q$_0$=0.5 throughout.
  
\section{H$_2$O Maser Observations and Results }

The Effelsberg 100-m telescope\footnote{The 100-m telescope at Effelsberg
is operated by the Max-Planck-Institut f{\"u}r Radioastronomie
in Bonn.} of the
MPIfR, equipped with a dual channel K-band HEMT receiver, was 
used to observe five sources known to have high columns of \HI\ 
in their central parsecs.  The system
temperature was 150--250 K on a main beam brightness temperature
scale. The beam size of the telescope at 22 GHz is
$\sim$40\arcsec. The data were obtained with an autocorrelator with
8$\times$512 channels and a bandwidth of 80 MHz (except on July 6,
2000; see below) for each of the eight backends, leading to channel
spacings of 4-5 \kms.  The measurements were carried out in a dual
beam switching mode with a switching frequency of 1 Hz and a beam
throw of 2\arcmin\ in azimuth. Pointing was checked every hour on
nearby continuum sources. Linear baselines were removed from the
summed spectra and we estimate the flux calibration to be accurate to
$\pm$15\%. Whenever not specifically mentioned, W3(OH) (3.2 Jy;
Mauersberger et al. 1988) was used for amplitude calibration.  No
\water\ emission was detected in any of the sources down to the rms
levels indicated below.  Upper limits for maser luminosities 
(with isotropic emission assumed here and throughout this paper) are given
in Table~1.
In some cases continuum detections are reported using the entire
0.5 GHz bandwidth available to the K-band HEMT receiver.


\subsection{Hydra~A}

Hydra~A is a high-luminosity FR I radio galaxy at a redshift of $z$ =
0.05384. \HI\ absorption with a width of 80 \kms\ is detected in this source 
and thought to lie in
a circumnuclear disk or torus with a scale height of $<$30 pc (Taylor
1996).  This source was observed at 21.09908 GHz on March 17,
2000. The continuum flux density at this frequency over a 0.5 GHz wide
band was 2.0$\pm$0.3 Jy.  The H$_2$O noise level obtained with 4.4
\kms\ channel spacing (the channel resolution is about 20\% wider) was
8 mJy for a $-$600 \kms $<$V$<+$580 \kms\ band relative to 21.09908
GHz.
 
\subsection{\pks}

\pks\ is the central cD galaxy in the cluster Abell 2597.  \HI\ is
detected on the kiloparsec scale (O'Dea \etal\ 1994), as well as on
smaller scales (Taylor \etal\ 1999).  On the parsec scale, two
distinct \HI\ absorption lines were detected, one narrow line (110
\kms\ FWHM) is seen toward the core and one of the inner jet
components, and a very broad line (735 \kms\ FWHM) appears toward only
the core.  This broad line is consistent with a compact, rapidly
rotating circumnuclear disk with a scale height of $<$20 pc (Taylor
\etal\ 1999).  This source was observed at 20.546 GHz ($z$ = 0.08220
for the H$_2$O rest frequency of 22.23508 GHz) on March 18, 2000. We
did not try to measure a continuum flux. Calibration at the sky
frequency was obtained towards NGC\,7027 (5.55 Jy according to Ott et
al. 1994). With 4.6 \kms\ channel separation we obtained an rms noise
level of 4 mJy for $-$1050 \kms $<$V$<+$1150 \kms\ relative to 20.546
GHz.

\subsection{NGC\,3894}

NGC\,3894 is a nearby ($z$=0.01068) elliptical galaxy with twin
relativistic parsec-scale jets (Taylor, Wrobel \& Vermeulen 1998).
The structure of the \HI\ detected in absorption in this source is
substantially more complicated than in Hydra~A or \pks. 
High resolution observations (Peck \&
Taylor 1998b) indicate that the absorbing gas might lie in a
circumnuclear torus, a large number of clouds along the line of sight
to the radio continuum source, or most likely, a combination of both.
This source was observed on 2000 March 19 and 2000 July 6, in the
latter case with 40 MHz backends. Rms limits are 4 mJy for 2800 \kms\
to 4300 \kms\ and 6 mJy for 2050 \kms\ to 2800 \kms\ LSR in March. In
July, the rms was 5 mJy for 2620 \kms\ to 3880 \kms. In both cases, the
channel separation was 4.2 \kms.

\subsection{NGC\,4151}

NGC\,4151 is a nearby ($z$=0.00332; distance=16.5 Mpc) Seyfert 1.5
galaxy which has been well-studied at most wavebands.  MERLIN
observations indicate that the \HI\ absorption in this source occurs
only toward the core with a linewidth of 90 \kms\
in the strongest component.  If this gas is in a toroidal structure,
it must be less than 50 pc in height (Mundell \etal\ 1995).  This
source was the nearest target in our sample.  NGC\,4151 was observed on
March 20, 2000. We obtained 3 mJy rms noise levels for 0 \kms\ to 2000
\kms\ LSR with a channel spacing of 4.2 \kms.

\subsection{1946+708}

1946+708 is a CSO at a redshift of $z$=0.101.  This source was
initially found to exhibit \HI\ absorption in VLA observations in 1995
(Peck, Taylor \& Conway 1999).  This absorption was then imaged at
very high angular resolution with the Global VLBI Network, showing the
region of broadest absorption (FWHM$\simeq$400 \kms) toward the core,
and several narrower components toward both the approaching and
receding jets.  The absorbing gas in 1946+708 is thought to occur
in a circumnuclear torus with an outer radius of at least 80 pc and a
scale height of $<$10 pc.  Further evidence for a torus in this source
is found in free-free absorption (Peck \& Taylor 2001).  This source
was observed with the Effelsberg 100-m telescope three times, on March
19, April 13, and July 6. As with NGC\,3894, the July data were
obtained with 40 MHz backends. The continuum flux was measured on
April 13: 0.27$\pm$0.04 Jy at 20.19535 GHz (with a bandwidth of 0.5 GHz).  
With a channel separation of 4.6 \kms\ we obtained 2-3 mJy rms
values between $-$1050 $<$ V $<$ 1180 \kms\ in March, 5 mJy between
$-$900 $<$ V $<$1000 \kms\ in April, and 4 mJy between $-$550$<$ V $<$
730 \kms\ in July (all velocities with respect to the central
observing frequency of 20.19535~GHz).


\section{\txs\ \HI\ Observations and Results}

The observations were made with Very Large Array\footnote{The National
Radio Astronomy Observatory is operated by Associated Universities,
Inc., under cooperative agreement with the National Science
Foundation.} at a center frequency of 1385 MHz on 2001 April 19 and
20.  The array was in the 'B' configuration, which provided an angular
resolution of $\sim$6\arcsec.  A total of 3.6 hours were obtained on
source on April 19th using 32 channels across a 6.25 MHz band to
provide a resolution of 43 km/s.  Both right and left circular
polarizations were observed.  Phase and bandpass calibration was
obtained by short (1 min) observations of the strong (2.1 Jy)
calibrator J2246{\tt -}1206 every $\sim$20 minutes.  The data were
loaded in real time and a rough reduction revealed the \HI\ detection.
For the observations on the following day the velocity resolution was
doubled (to 22 km/s) at the expense of only recording in right
circular polarization.  A total of 2.3 hours were obtained on source.

Following standard VLA calibration practices, a continuum visibility
data set and image were generated from the line data for each epoch by
averaging the line-free channels.  These data were then
self-calibrated to remove small residual phase errors.  The rms noise
in the final continuum image of \txs\ is 0.10 mJy.  The continuum
emission from \txs\ is unresolved (size $<$3 kpc) and
there are no other sources stronger than 0.3 mJy within one arcminute.
The self-calibration solutions were then transferred to the line data.
Spectral line cubes were made for \txs\ and all the sources in the
field (see Fig.~1) stronger than 2 mJy simultaneously using the AIPS
task IMAGR.  The rms noise in a single channel from the combined data
set is 0.25 mJy/beam, while in the higher spectral resolution data the
rms noise in a channel is 0.66 mJy/beam.  The higher noise is the
result both narrower channels (by factor 0.5) and less data (factor
0.24).

In Fig.~2 we present the 43 km/s resolution spectrum of \txs.  In
addition to a component with width $\sim$125 km/s, there appear
substantial wings to the line, indicative of a broader (420 km/s) 
component.  Two
Gaussian components have been fit to the data and are drawn
individually in Fig.~2 along with their sum.  Parameters of the fits
are given in Table~2.  The higher resolution (22 km/s) spectrum is
shown in Fig.~3 with the Gaussian fits overlaid.

\subsection{\txs\ Continuum Observations and Results}

The continuum observations were made with Very Large Array on 2001
April 24 at C, X, U, and K bands.  A total of 1 hour was used to
obtain the measurements presented in Fig.~4 and Table~3.  The
observations at frequencies of 15 GHz and below consisted of a single
scan on \txs\ of 1--3 minutes duration bracketed by a 1 minute
integration on the strong calibrator J2246{\tt -}1206.  At 22 GHz we
employed the fast switching mode using a 3 minute cycle time for
18 minutes and the
nearby calibrator J2236{\tt -}1433.  From 1.4 to 5 GHz the 
spectral index, $\alpha$, where $S_\nu \propto \nu^\alpha$,
is $\alpha = -0.66$.  Between 5 and 22 GHz the 
spectrum steepens to a power law of slope $-$0.92.  

The highest resolution, obtained at 22 GHz, is 0.39 $\times$ 0.21\arcsec\ 
in position angle $-$12\deg.  At this resolution \txs\ is found
to be slightly resolved with an extension along PA 145.3\deg.  This
is in good agreement with the 'A' configuration VLA image presented by 
Falcke \etal\ (2000) at 8.46 GHz.

\section{Discussion}

\txs\ contains the most
luminous known H$_2$O megamaser with a luminosity in the 1.3 cm line of
6100 $L_\odot$ (Koekemoer et al.\ 1995). The water maser emission from
\txs\ is fairly broad, with a FWHM of 88 km s$^{-1}$.  Recent HST
observations by Falcke et al.\ (2000) classify the galaxy as a spiral
and reveal a dust lane cutting across the nucleus.  Falcke et al.\
(2000) also present VLA observations at 8.4 GHz showing that the radio
emission is compact ($<$ 1\arcsec), symmetric, and has an axis
perpendicular to the dust lane.  No larger-scale diffuse emission is
present to the sensitivity limits of the Northern VLA Sky Survey
(NVSS) at about 2 mJy/beam (Fig.~1; Condon et al.\ 1998).

Although the $\sim$420 \kms\ absorption probably results from 
neutral material associated with the
atomic and molecular torus thought to feed the active nucleus,
the deeper $\sim$125 \kms\ line in
\txs\ could be indicative of an 
interaction between the radio jet and surrounding material.  The water
maser emission is also fairly broad (88 km/s) in \txs, and could
likewise originate from the central torus or from a jet-cloud
interaction.  VLBA observations of both the \water\ emission and \HI\
absorption could help to discriminate between these two models.

Both the \water\ megamaser emission and the \HI\ absorption are
thought to arise from the central 0.1 to 100 parsecs of the molecular
and atomic torus around an AGN.  In searching for these two processes,
however, it is important to keep in mind that the observed flux
density of the maser emission falls off with distance (although this
may be less steep than the inverse square-law if the water maser
emission is amplifying a beamed nuclear continuum), while the \HI\
absorption relies on the presence of suitably bright radio emission,
but is independent of distance.  Perhaps because of this difference,
maser emission and \HI\ absorption studies thus far have focused on
different classes of AGN.  The majority of megamaser surveys have
targeted low luminosity AGN, predominantly optically selected and
relatively nearby Seyferts and LINER galaxies, while the \HI\ surveys,
encompassing a much smaller number of sources but usually obtaining
longer integrations, have mainly targeted powerful radio galaxies.
There has been one megamaser survey directed toward FRI radio galaxies
(Henkel et al. 1998) and a survey of FRII galaxies and BL Lac objects
has also been made (E. Ros, A. Tarchi, priv.  communication), but no
detections were found and the upper limits of the two latter studies
have yet to be determined.


In Table~1 we summarize the properties of all known \water\ megamasers
and the \HI\ absorption detections in these systems to date.  This
table illustrates a fair amount of overlap with 8/19 (42\%) of the
sources having both \water\ megamaser emission and \HI\ absorption.
Not all megamaser sources shown have been searched carefully for \HI\ 
absorption, so the actual overlap could be somewhat larger. 

Given the fairly high incidence of \HI\ absorption in systems with
\water\ megamaser emission previously reported in the literature, we
carried out a moderately deep search for \water\ megamaser emission in
five sources with optical depths in \HI\ of 0.1 or more.  None of
these sources were detected (see Table~1 for the luminosity limits
derived from the observations).  The luminosity limits for the two
closest sources, NGC\,4151 and NGC\,3894 are below those of all known
\water\ megamasers, so in the case of isotropic emission we would have
expected to detect those sources (see Fig.~5).  For the other,
more-distant, systems the luminosity limits are fairly high so it is
still possible that faint \water\ megamasers reside there, or that
these systems are also beamed away from us.  As evident from Fig.~5,
there is only a weak correlation between radio power and \water\ maser
luminosity, with a Spearman's rank-order correlation coefficient of
0.465 leading to a probability that the two quantities are correlated
at the 97.5\% level.  The \HI\ absorbers detected so far appear to be
distributed in a similar way as the megamasers without any \HI\
absorption.

There are several possible reasons for not detecting \water\ emission
from the nearby sources, even if the interpretation of the \HI\ is
correct, and a pc scale molecular disk exists within the torus, as in
the scenario proposed by Neufeld \& Maloney (1995) based on the
prototypical source NGC~4258. First, the molecular disk is expected to
be extremely thin and warped by radiation pressure (Pringle 1996),
making the detection of \water\ much more dependent on the geometry
and orientation of the disk than the detection of \HI\ absorption is
on the orientation of the larger torus.  In some sources, the radio
axis might be close to the plane of the sky, but the regions in the
warped disk which have suitable temperatures and pressures to obtain
population inversion do not lie along our line of sight to a continuum
source nor have a long enough path length to generate
self-amplification.  Another possibility is that the model wherein a
molecular disk is embedded in the \HI\ torus simply does not scale to
higher powered AGN.  A higher accretion rate, a stronger gravitational
potential, a stronger central x-ray source, or a combination of these
factors might alter the temperature and pressure of the mid-plane of
the torus and thus the conditions for establishing \water\ masers is
not met in these systems.  Conversely, the non-detection of \HI\
absorption in a few megamaser galaxies might indicate that the \water\
emission is the result of an interaction between the jet and a nearby
molecular cloud, rather than arising in an accretion disk, as is
thought to be the case in Mrk~348 (Peck \etal\ 2001, in preparation),
or could possibly be the result of very high spin temperatures in the
atomic part of the torus, yielding extremely low \HI\ optical depths.

Braatz, Wilson \& Henkel (1997) find a detection rate for
extragalactic \water\ masers of 7.2\% in Seyferts 2's and LINERs in a
distance-limited sample.  This is a greater success rate than we
achieved in selecting systems with \HI\ absorption, but our sample is
still too small to say anything meaningful about the difference.
However, the number of sources with known \HI\ absorption is
increasing dramatically thanks to new facilities at the Giant
Meterwave Radio Telescope (GMRT) and Westerbork Synthesis Radio
Telescope (WSRT).  The detection of a torus signature in either \HI\
or \water\ masers is of interest since it can then be followed up with
VLBA studies which can provide important information about the spatial
distribution and velocity field of the accreting gas and the mass of
the central engine.


\section{Conclusions}

We report on the detection of broad \HI\ absorption in the 
gigamaser galaxy \txs.  Sensitive VLBI observations are 
being planned in order to investigate the location and 
kinematics of the \HI\ gas.  It will also be of interest to
learn more about the parsec-scale morphology of this 
exceptional object.

Among known \water\ masers the incidence of \HI\ absorption
is fairly common ($>$42\%).  Although a moderately deep 
survey of five radio galaxies failed to yield any new 
\water\ megamasers, there is still some promise to a 
physically motivated search.  In the near future it should
be possible to start with larger samples of \HI\ absorbing
systems and refine our understanding of the connection
between the \HI\ absorbers and the \water\ emitters.

\acknowledgments
We thank the referee for insightful comments on an 
earlier draft of the manuscript.
This research has made use of the NASA/IPAC Extragalactic Database (NED)
which is operated by the Jet Propulsion Laboratory, Caltech, under
contract with NASA. This research has also made use of NASA's Astrophysics
Data System Abstract Service.

\clearpage

\tightenlines
\singlespace

\begin{center}
\begin{deluxetable}{l c c c c c r r c c c c}
\tablecolumns{10}
\tablewidth{-10pc}
\tablecaption{Megamaser and \HI\  Detection Summary}
\tabletypesize{\scriptsize}
\tablehead{
& Host & Isotropic & H$_2$O &H$_2$O &Maser & Isotropic &  Peak \HI\ & Peak \HI\ &\HI\  \\
Megamaser  & Galaxy &  Lum. & FWHM &V$_{\rm LSR}$& Ref. & $P_{\rm 1.4 GHz}$ & Opt. Depth &FWHM & ref. \\
Source  & $\bullet$ &  (L$_{\odot}$) & (\kms) &(\kms) &  & (W Hz$^{-1}$) & ($\tau$) &(\kms) &}
\startdata
Circinus & Sy2 & 24 & 1& 274--562& 1 & $6.8 \times 10^{21}$ & \nodata & $\sim$180 & 2 \\
ESO~103-G35 & Sy1/2 & 360 &8&4073& 1 & \nodata &\nodata&\nodata&\nodata\\
IC~2560 & Sy2 & 130 &25&2919 & 1 & $5.6 \times 10^{21}$    &\nodata&\nodata&\nodata \\
IC~1481 & L & 320 &2&6235--6237 & 1 & $2.9 \times 10^{22}$ &\nodata&\nodata&\nodata  \\
Mrk~1 & Sy2 & 64 & 3 -- 6&4868--4869 & 1 & $3.8 \times 10^{22}$  &\nodata&\nodata&\nodata\\
Mrk~348 & Sy2 & 450 & 133&4610--4678& 3 & $1.4 \times 10^{23}$ &$<$0.004& & 4\\
Mrk~1210 & Sy2 & 99 &0.8 -- 11& 4213--4314& 1 & $4.2 \times 10^{22}$ &\nodata&\nodata&\nodata\\
NGC\,1052 & L/Sy2 & 140 &80 -- 120&1605--1668& 1 & $4.0 \times 10^{22}$ & 0.021 & 35 & 5\\
NGC\,1068 & Sy1/2 & 170& &800--1500& 1 & $1.4 \times 10^{23}$ &0.078 & 161 & 4\\
NGC\,1386 & Sy2 & 120 &4 &969& 1 & $6.1 \times 10^{20}$ &\nodata&\nodata&\nodata\\
NGC\,2639 & Sy1.9 & 71 &1 -- 7 &3300--3323& 1 & $2.6 \times 10^{22}$ & $<$0.003 & & 4\\ 
NGC\,3079 & L/Sy2 & 520 &10 &926--1190& 1,6 & $2.3 \times 10^{22}$ &0.8 & 133 & 4,6\\
NGC\,3735 & Sy2 & 12 & 1.4--3&2262--2268& 7 & $1.3 \times 10^{22}$ &\nodata&\nodata&\nodata\\
NGC\,4258 & L/Sy1.9 &85& 1 -- 7 &-365--1460& 1,8 & $3.6 \times 10^{21}$ &$<$1.4&\nodata& 9 \\
NGC\,4945 & Sy2 & 57 &3& 694& 1 & $4.4 \times 10^{22}$ &0.29&$\sim$250&10 \\ 
NGC\,5347 & Sy2 & 32 &15 &2376--2401& 1 & $6.5 \times 10^{20}$ & \nodata&\nodata&\nodata\\
NGC\,5506 & Sy1.9 & 61 &0.7 -- 6&1731--1821 & 1 & $2.5 \times 10^{21}$ &0.109 & 95 & 4\\
NGC\,5793 & Sy2 & 125 &1.3, 14 &3190--3677 &11 & $3.0 \times 10^{23}$ &3& 16, 50 & 12\\
TXS~2226{\tt -}184 & E/S0 & 6100 &88 & 7570 & 1 & $8.7 \times 10^{22}$ &0.1& 125, 418& 13\\
\tableline
& Host &  3$\sigma$ H$_2$O &Lum.  &V$_{\rm LSR}$ &Maser & Isotropic & Peak \HI\ & Peak \HI\ &\HI\\\
\HI\ Torus & Galaxy & Upper lim. &limit\tablenotemark{\dag} &Searched\tablenotemark{*} &Ref. & $P_{\rm 1.4 GHz}$ & Opt. Depth & FWHM & ref.  \\
Source  & & (mJy) &(L$_{\odot}$) & (\kms) & & (W Hz$^{-1}$) & ($\tau$) &(\kms) &   \\
\tableline
Hydra~A & E & 13& $<$140 &15552--16732 &13 & $2.3 \times 10^{26}$  &0.99 &80 &14 \\
PKS~2322{\tt -}123& E & 12 & $<$310 &23610--25811 &13 & $2.5 \times 10^{25}$ &0.3,0.6 &735, 110 &15 \\
NGC\,3894& E & 12 & $<$ 5  &2050--4300 &13 & $8.7 \times 10^{22}$ &0.1 &80 &16 \\
NGC\,4151& Sy1.5 & 9 & $<$0.6 &0--2000 &13 & $7.0 \times 10^{21}$ &0.23 &90 &17 \\
1946+708& E & 12 & $<$470 &29250--31480 &13 & $2.0 \times 10^{25}$ &0.1 &375 &18 \\
\tableline   
\enddata

\tablecomments{References: (1) Braatz et al 1996 and references therein, 
(2) Koribalski 1996, (3) Falcke \etal\ 2000, (4) Gallimore et al 1999, 
(5) van Gorkom, \etal\ 1986, 
(6) Sawada-Satoh \etal\ 2000, (7) Greenhill \etal\ 1997, 
(8) Greenhill \etal\ 1995, (9) Mundell \etal\, in preparation,
(10) Ott \etal\ 2001,
(11) Hagiwara \etal\ 2000, 
(12) Pihlstr\"om \etal\ 2000, (13) this paper, (14) Taylor 1996, 
(15) Taylor \etal\ 1999, (16) Peck \& Taylor 1998b, (17) Mundell \etal\ 1995, 
(18) Peck, Taylor \& Conway 1999 \\
\noindent
$^{\bullet}$ Sy=Seyfert, L=LINER, E=Elliptical. \\
\noindent
$^{\dag}$ calculated assuming a linewidth of 10 \kms. \\
\noindent
$^{*}$ optical velocity convention (cz).}
\end{deluxetable}
\end{center}

\clearpage
\begin{center}
\tightenlines
\singlespace
TABLE 2 \\
\smallskip
\txs\  \HI\ C{\sc olumn} D{\sc ensities}
\smallskip
 
\begin{tabular}{l r r r r r r r r}
\hline
\hline
& $S_{\rm 1.4 GHz}$ &  $S_{\rm line}$ & $V$ & $\Delta V$ & $\tau$ & $N_{\rm H}$ \\
Component  & (mJy) & (mJy) & (km s$^{-1}$) & (km s$^{-1}$) & & (cm$^{-2}$) \\
(1) & (2) & (3) & (4) & (5) & (6) & (7) \\
\hline
\noalign{\vskip2pt}
narrow & 73.5 & 12.3 $\pm$ 0.98 & 7491 $\pm$ 3 & 125 $\pm$ 10  & 0.183 & 3.3 $\times 10^{23}$ \\
broad & 73.5 & 5.6 $\pm$ 0.74 & 7454 $\pm$ 18 & 418 $\pm$ 72 & 0.079 & 4.8 $\times 10^{23}$ \\
\hline
\end{tabular}
\end{center}
\begin{center}
{\sc Notes to Table 2}
\end{center}
\noindent
Col.(1).---Component name.
Col.(2).---Continuum flux density at 1.4 GHz in mJy.
Col.(3).---Depth of the line in mJy.
Col.(4).---Central velocity (c$z$) in km s$^{-1}$.
Col.(5).---FWHM in velocity in km s$^{-1}$.
Col.(6).---Optical depth, assuming a continuum source
covering factor of unity.
Col.(7).---The column density in units of cm$^{-2}$ calculated assuming
a spin temperature of 8000 K (Conway \&
Blanco 1995). If the \HI\ absorption results from gas outside the
nuclear region then a more reasonable assumption for the
spin temperature would be 100 K, reducing the column densities quoted
here by a factor of 80.  
\bigskip

\begin{center}
TABLE 3 \\
\smallskip
C{\sc ontinuum} O{\sc bservations}
\smallskip
 
\begin{tabular}{r r r r }
\hline
\hline
$\nu$ & $S_{\rm VLA}$ & time & $\Delta\nu$ \\
(GHz) & (mJy) & (min) & (MHz) \\
\hline
\noalign{\vskip2pt}
 1.385  &          73.3 $\pm$ 2.2\phantom{0} & 350\phantom{.0} & 5 \\
 4.860  &          32.2 $\pm$ 0.98 & 1.4 & 50 \\
 8.460  &          20.0 $\pm$ 0.60 & 0.8 & 50 \\
14.940  &          11.7 $\pm$ 0.38 & 2.0 & 50 \\
22.460  &           7.87 $\pm$ 0.41 & 10.2 & 50 \\ 
\hline
\end{tabular}
\end{center}
\smallskip

\clearpage

\bigskip
\clearpage

\begin{figure}
\plotone{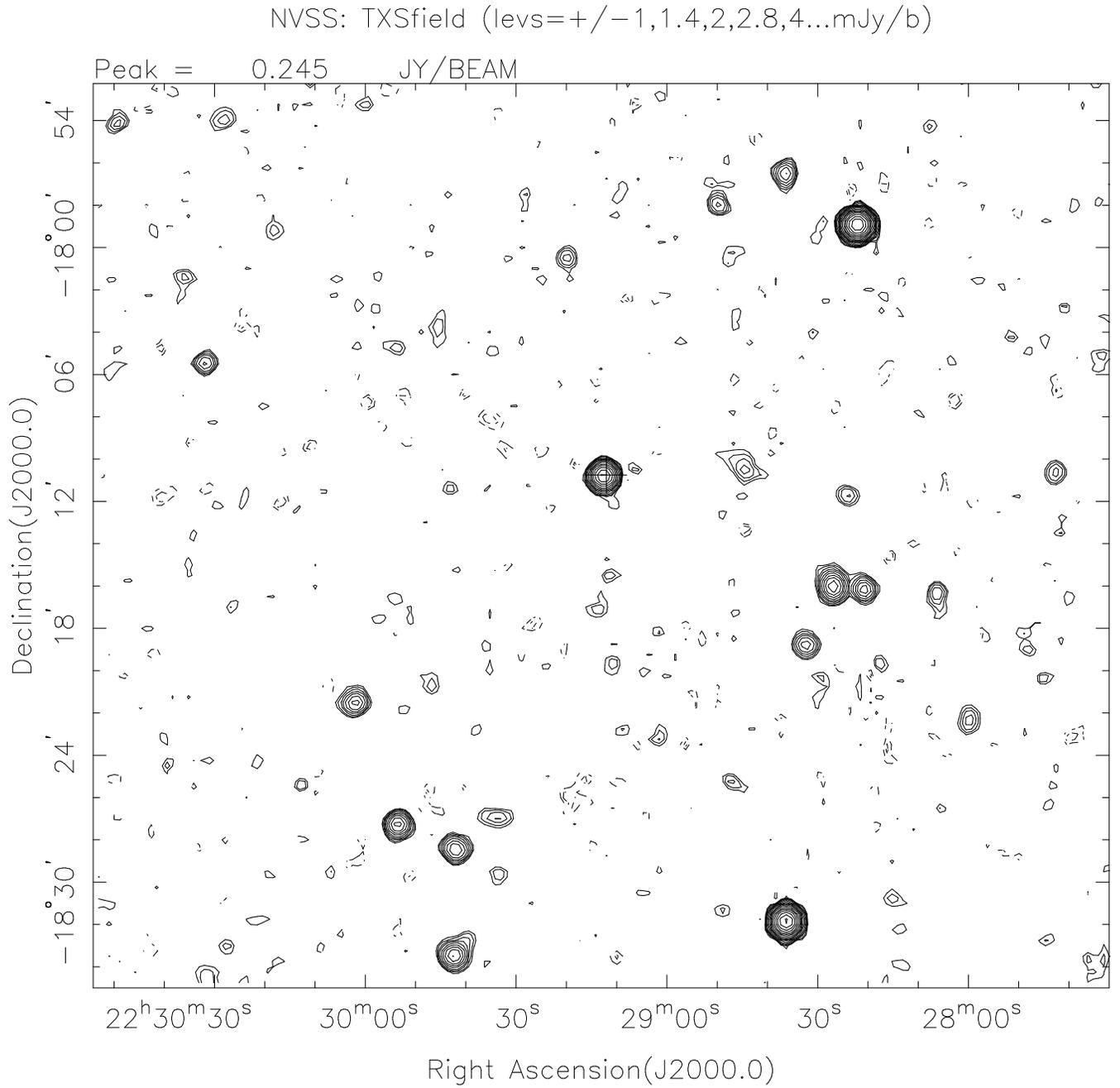}
\caption{The NVSS (Northern VLA Sky Survey) image from Condon et al.\ 1998.  Spatial resolution is
45\arcsec.  Contours are drawn at $-$1, 1, 1.4, 2, ..., 181 mJy/beam 
by $\sqrt{2}$ intervals with 
negative contours shown dashed. The cross marks the position of \txs.
\label{fig1}}
\end{figure}

\begin{figure}
\plotone{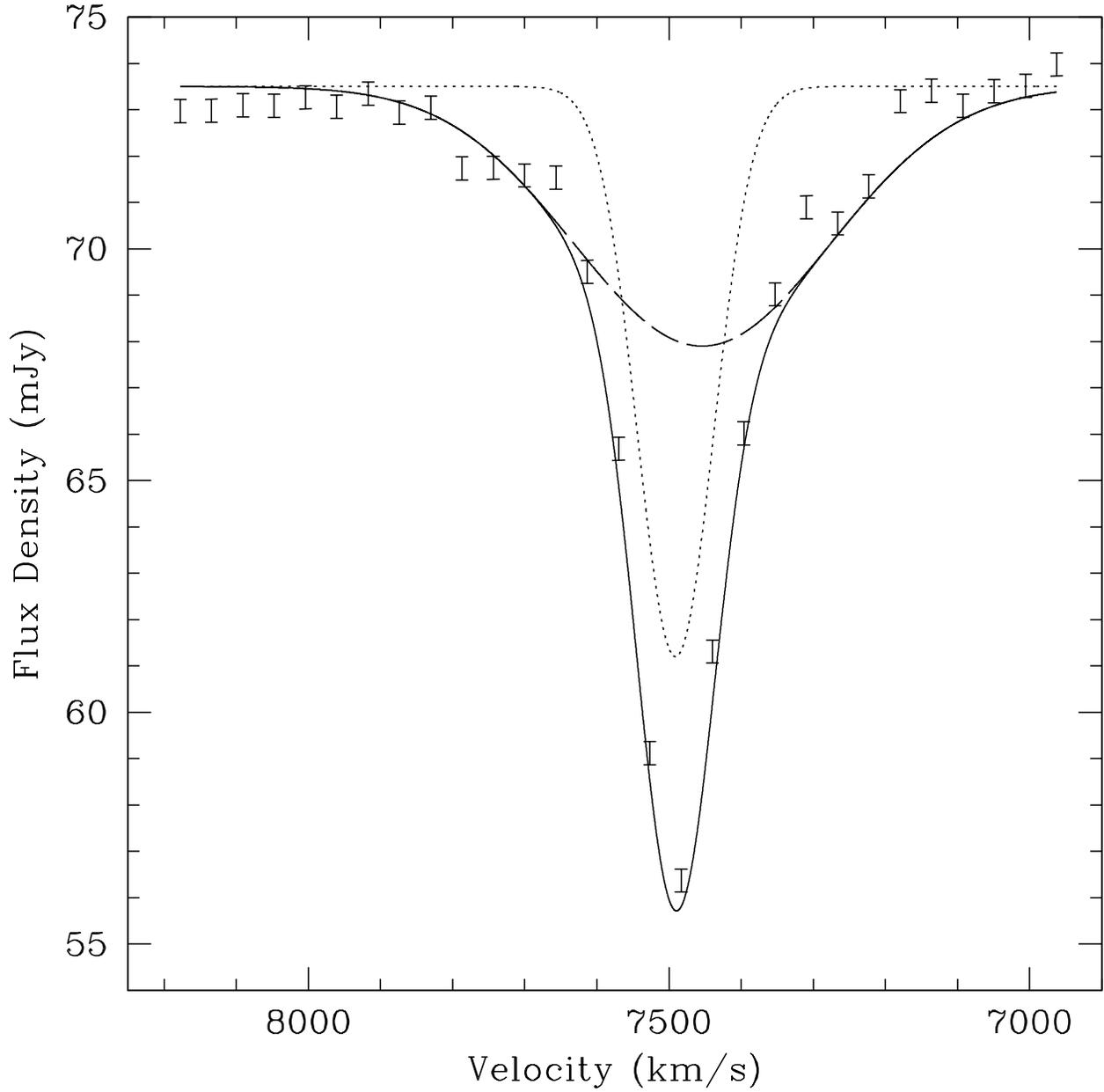}
\caption{The low resolution \HI\ spectrum towards \txs\ derived from 
combined observations on 2001 April 19 and 20.  The velocity frame
is LSR (Local Standard of Rest) and the velocity  
resolution is 43 \kms.  The solid line represents the best fit
of two Gaussians (dotted and long-dash lines) to the data. 
\label{fig2}}
\end{figure}

\begin{figure}
\plotone{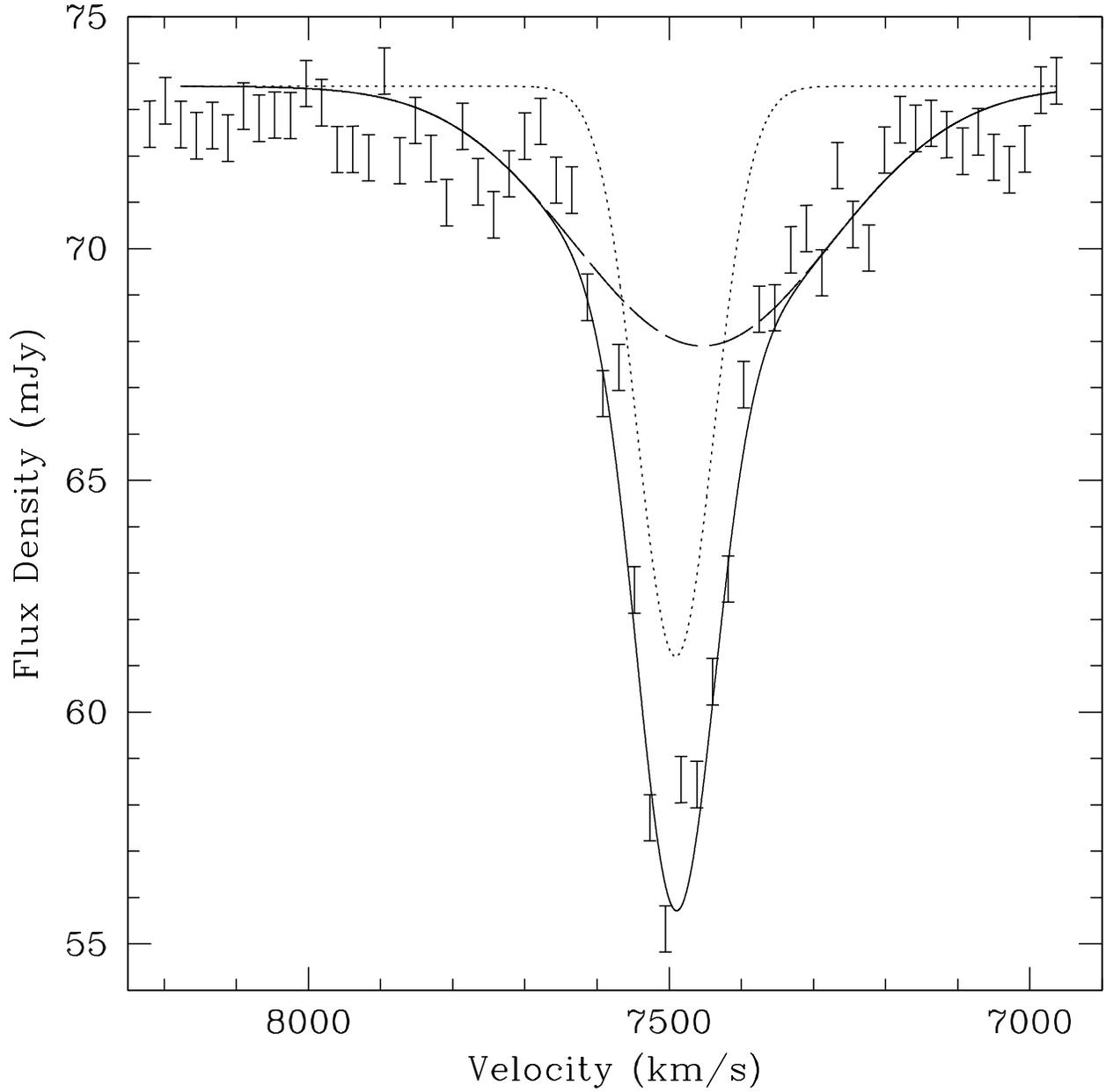}
\caption{The high resolution \HI\ spectrum towards \txs\ derived from 
observations on 2001 April 20.  The velocity frame used is LSR and the
velocity resolution is 22 \kms.   The solid line represents the best fit
of two Gaussians (dotted and long-dash lines) to the data. 
\label{fig3}}
\end{figure}

\begin{figure}
\plotone{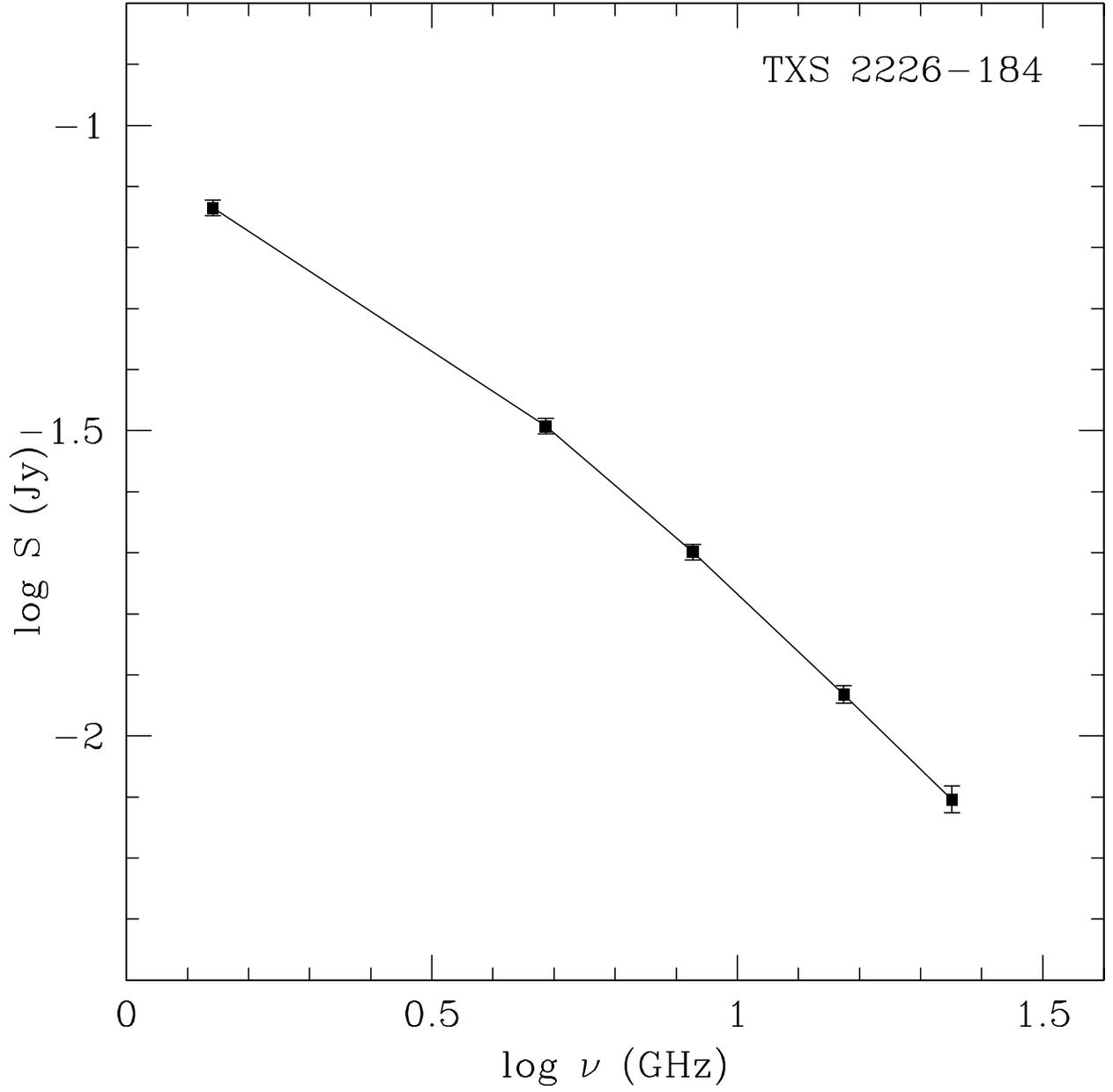}
\caption{The broadband continuum emission spectrum towards \txs\ derived from 
observations in April 2001.
\label{fig4}}
\end{figure}

\begin{figure}
\plotone{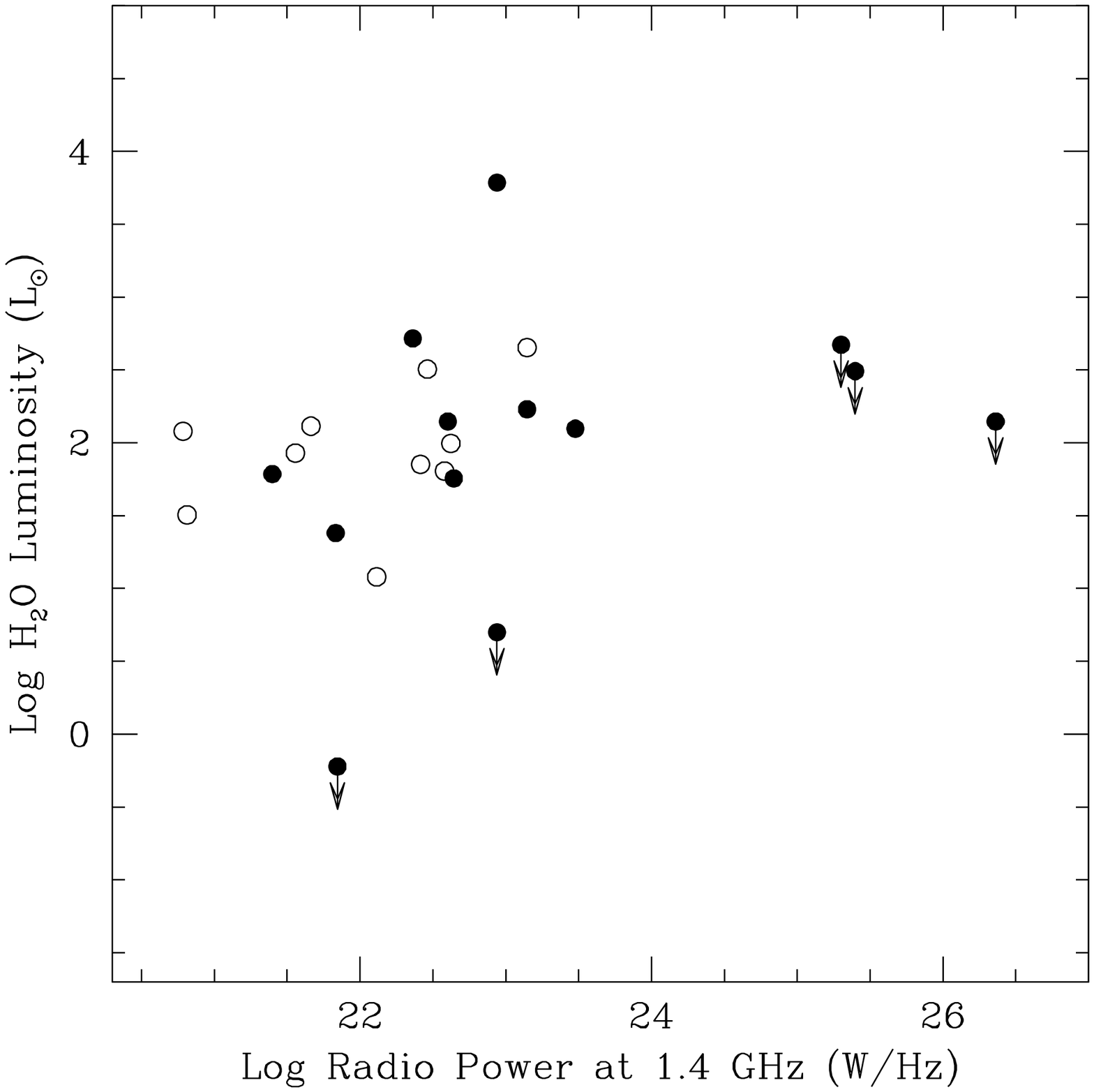}
\caption{A plot of \water\ Maser Luminosity against the Radio Power at 
an observed frequency of 1.4 GHz for all the sources in Table~1.  
Sources plotted with filled circles
have \HI\ detected in absorption, while those plotted with empty 
circles have either not been detected or not been observed.
\label{fig5}}
\end{figure}


\clearpage

\begin{references}

\reference{bra97}Braatz, J. A., Wilson, A. S. \& Henkel, C. 1997, ApJS, 
110, 321

\reference{bra96}Braatz, J. A., Wilson, A. S. \& Henkel, C. 1996, ApJS, 
106, 51

\reference{con98}Condon, J.J., Cotton, W.D., Greisen, E.W., Yin, Q.F., Perley,
R.A., Taylor, G.B., \& Broderick, J.J.\ 1998, AJ, 115, 1693

\reference{con96}Conway, J. E. 1996 in {\it The Second Workshop on Gigahertz
Peaked Spectrum and Compact Steep Spectrum Radio Sources},
ed. I. A. G. Snellen, R. T. Schilizzi, H. J. A. Rottgering and
M. N. Bremer [Leiden Observatory], 198

\reference{con95}Conway, J. E \& Blanco, P. R. 1995 ApJ, 449, L131

\reference{fal00}Falcke, H., Wilson, A.S., Henkel, C., Brunthaler, A., 
\& Braatz, J.A.\ 2000, ApJ, 530, L13

\reference{gal96}Gallimore, J. F., Baum, S. A. and O'Dea, C. P. 1996 ApJ, 
464, 198

\reference{gal99}Gallimore, J. F., Baum, S. A., O'Dea, C. P., Pedlar,
A. \& Brinks, E. 1999, ApJ, 524, 684

\reference{jvg86}van Gorkom, J. H., Knapp, G. R., Raimond, E., Faber, S. M.,
\& Gallagher, J. S.\ 1986, AJ, 91, 791

\reference{jvg89}van Gorkom, J. H., Knapp, G. R., Ekers, R. D., Ekers, D. D.,
Laing, R. A., \& Polk, K. S. 1989, AJ, 97, 708

\reference{gre97}Greenhill, L. J. \& Gwinn, C. R. 1997, Ap\&SS, 248, 261

\reference{gr297}Greenhill, L. J., Herrnstein, J. R., Moran, J. M., Menten, K. M. \& Velusamy, T. 1997, ApJ, 486, L15

\reference{gre95}Greenhill, L. J., Henkel, C., Becker, R., Wilson, T. L. \& Wouterloot, J. G. A. 1995, A\&A, 304, 21

\reference{hag00}Hagiwara Y., Diamond, P. J., Nakai, N. 2000, in {\it
Proceedings of the 5th EVN Symposium} Eds. J. Conway, A. Polatidis,
R.Booth \& Y. Pihlstr\"om, Onsala Space Observatory, Chalmers
Technical University, Gothenburg, Sweden

\reference{hen98}Henkel, C., Wang, Y.~P., Falcke, H., Wilson,
A.~S. \& Braatz, J.~A. 1998, A\&A, 335, 463

\reference{koe95}Koekemoer, A. M., Henkel, C., Greenhill, L.J., Dey,
A., van Breugel, W., Codella, C., \& Antonucci, R.\ 1995, Nature, 378,
697

\reference{kor96}Koribalski B., 1996, in {\it The Minnesota Lectures
  on Extragalactic Neutral Hydrogen} ASP Conf. Ser. Vol. 106,
  ed. Skillman E. D., [ASP: San Francisco], p.238 
  

\reference{mau88}Mauersberger, R., Wilson, T. L. \& Henkel, C. 1988,
A\&A 201, 123

\reference{miy95}Miyoshi, M., Moran, J., Herrnstein, J., Greenhill, L.,
Nakai, N., Diamond, P. \& Inoue, M. 1995, Nature, 373, 127

\reference{mun95}Mundell, C. G., Pedlar, A., Baum, S. A., O'Dea,
C. P., Gallimore, J. F., \& Brinks, E. 1995, MNRAS, 272, 355

\reference{neu95} Neufeld, D. A. \& Maloney, P. R. 1995, ApJ, 447, L17


1994, ApJ, 436, 669

\reference{ott01}Ott, M., Whiteoak. J. B., Henkel, C. \& Wielebinski, R. 2001, A\&A, 372, 463 

\reference{ott94}Ott, M., Witzel, A., Quirrenbach, A., Krichbaum, T. P., Standke, K. J., Schalinski, C. J. \& Hummel, C. A. 1994, A\&A 284, 331

\reference{pec98}Peck, A. B. \& Taylor, G. B. 1998a, BAAS, 193, 620

\reference{pec98}Peck, A. B., \& Taylor, G. B. 1998b, ApJ, 502, L23


\reference{pec99}Peck, A. B., Taylor, G. B. \& Conway, J. E. 1999, ApJ, 
521, 103

\reference{pec98}Peck, A. B., \& Taylor, G. B. 2001, ApJ, 554, L147

\reference{pih00} Pihlstr\"om, Y.M., Conway, J.E., Booth, R.S., Diamond, P.J. \&Koribalski, B.S. 2000 A\&A, 357, 7

\reference{pih01} Pihlstr\"om, Y.M. 2001, Ph.D. Thesis, Chalmers University

\reference{pri96}Pringle, J. E. 1996, MNRAS, 281, 357

\reference{sat00}Sawada-Satoh, S., Inoue, M., Shibata, K. M., Kameno, S., Migenes, V., Nakai, N. \& Diamond, P. J. 2000, PASJ, 52, 421

\reference{sat97}Satoh, S., Inoue, M., Nakai, N., Shibata, K. M.,
Kameno, S., Migenes, V., Diamond, P. J. 1997, in {\it The Central
Regions of the Galaxy and Galaxies}, IAU Symposium 184, 208

\reference{tay96}Taylor, G. B. 1996, ApJ, 470, 394

\reference{tay98}Taylor, G. B., Wrobel, J. M., \& Vermeulen, R. C.\ 1998, 
ApJ, 498, 619 

\reference{tay99}Taylor, G. B., O'Dea, C. P., Peck, A. B.\ 
\& Koekemoer, A. M. 1999, ApJ, 512, L27   


\end{references}
\end{document}